\documentclass{PoS}

\title{{\itshape XMM-Newton} discovery of a possible cyclotron emission feature from the SFXT IGR J18483$-$0311}

\ShortTitle{Possible cyclotron emission feature from the SFXT IGR J18483$-$0311}

\author{\speaker{V. Sguera}\thanks{The authors acknowledge the ASI
financial support via grant ASI-INAF  I/033/10/0 and I/009/10/0.}\\
        INAF, Istituto di Astrofisica Spaziale e Fisica Cosmica, Via Gobetti 101, I-40129 Bologna, Italy\\
        E-mail: \email{sguera@iasfbo.inaf.it}} 

\author{L. Sidoli\\
       INAF, Istituto di Astrofisica Spaziale e Fisica Cosmica, Via E. Bassini 15, I-20133 Milano, Italy\\
}
\author{A. Bazzano\\
       INAF, Istituto di Astrofisica Spaziale e Fisica Cosmica, Via Fosso del Cavaliere 100, I-00133 Rome, Italy\\
}
\author{L. Bassani\\
INAF, Istituto di Astrofisica Spaziale e Fisica Cosmica, Via Gobetti 101, I-40129 Bologna, Italy \\
}
\author{M. Orlandini\\
INAF, Istituto di Astrofisica Spaziale e Fisica Cosmica, Via Gobetti 101, I-40129 Bologna, Italy \\
}

\abstract{We report the results from an archival {\itshape XMM-Newton} observation of the Supergiant Fast X-ray Transient (SFXT)
IGR J18483$-$0311 during its apastron passage.  The measured  0.5--10 keV luminosity state (1.3$\times$10$^{33}$ erg   s$^{-1}$) 
is the lowest ever reported in the literature, it is best  fitted  by an absorbed black body model yielding
parameters consistent with previous measurements. In addition, we find evidence
of an emission line feature at $\sim$3.3 keV in  the 0.5--10 keV EPIC-pn source spectrum. 
We show that its physical explanation in terms of atomic emission line appears unlikely and 
conversely we attempt to ascribe  it to an electron cyclotron emission line which would implies  
a neutron star magnetic field of the order of $\sim$3$\times$10$^{11}$ G. A possible hint 
of the first harmonic  is also found.
If firmly confirmed by future longer X-ray observations, this would be the first detection ever 
of a cyclotron feature in the X-ray spectrum of a SFXT, with
important implications on theoretical models.}

\FullConference{8th INTEGRAL Workshop The Restless Gamma-ray Universe - INTEGRAL 2010,\\
		September 27-30, 2010\\
		Dublin Ireland}

\begin{document}

\section{Introduction}

During the last few years, the INTEGRAL  satellite  has played a key role in discovering a previously unrecognized class of high-mass X-ray binaries which display a very unusual and intriguing fast X-ray transient behaviour and host a massive OB supergiant star as companion donor: the so-called Supergiant Fast X-ray Transients (SFXTs; Sguera et al. 2005, 2006; Negueruela et al. 2006). To date $\sim$10 firm SFXTs are known (plus the same number of candidates), they are characterized by long periods of low X-ray activity level (with luminosities values or upper limits in the range 10$^{32}$--10$^{34}$ erg s$^{-1}$), interspersed with short, strong flares lasting from a few hours to no more than a few days and reaching peak luminosities of  10$^{36}$--10$^{37}$ erg s$^{-1}$. The broad band X-ray spectra of SFXTs are very similar to those of accreting X-ray pulsars (i.e. flat power law below 10 keV and cut-off at 10--30 keV) strongly suggesting that they host a neutron star as well. Indeed X-ray pulsations 
have been detected in four SFXTs with spin periods ranging from 5 to 230 seconds.  

The physical reasons behind the intriguing SFXTs behaviour are still not explained and  several theoretical mechanisms  have been proposed in the literature (see Sidoli 2009  for a review). 
All such models are based on the general consensus that  the supergiant wind is highly inhomogeneus and structured (clumpy),  however they can be mainly divided into  two major groups  i) models invoking spherically symmetric/anisotropic clumpy winds ii) models invoking gated mechanisms able to stop/allow 
the accretion onto a highly magnetized (B$\sim$10$^{14}$ G, i.e. magnetar)  and slow (P$\geq$1000 s) neutron star. 

IGR J18483$-$0311 is one of the firm known SFXT (Sguera et al. 2007). It is composed by a 21 seconds 
neutron star as compact object and a  B0.5Ia supergiant star as companion donor located at $\sim$3 kpc (Rahoui \& Chaty 2008).
To date several outbursts were observed  by INTEGRAL and Swift showing luminosities
and durations  up to  $\sim$3$\times$10$^{36}$ erg  s$^{-1}$  and $\sim$3 days, respectively.
The orbit of the system is likely small and eccentric  with a period of $\sim$18.5 days (Sguera et al. 2007, Romano et al. 2010).

Here we report the results from an archival {\itshape XMM-Newton} observation of IGR J18483$-$0311.

\section{{\itshape XMM-Newton} observation and results}
An {\itshape XMM-Newton} observation of the source, performed on 12 October 2006 
for a total exposure of $\sim$20 ks, was fortunately timed catching it right during the apastron passage at $\phi$=0.52.
However, after  rejecting time intervals  affected by high background the net good exposure time reduced to 14.4 ks. 
The source net count rates (1--10 keV) are 0.097$\pm{0.003}$ counts~s$^{-1}$ (pn), 0.035$\pm{0.002}$ 
counts~s$^{-1}$ (MOS1), and 0.027$\pm{0.001}$ counts~s$^{-1}$ (MOS2). 
Since the spectroscopy with the MOS cameras does not provide any improvement with
respect to the analysis of  the pn spectrum alone, in the following 
we report only on the source EPIC pn results.

We first fit the 0.5--10 keV  EPIC-pn spectrum of IGR J18483$-$0311  with an absorbed power law model
whose best fit parameters (N$_{H}$ = 7.8$^{+1.3}_{-1.1}$ $\times$10$^{22}$ cm$^{-2}$, $\Gamma$=2.4$\pm$0.3) are in fair agreement with 
those already reported by Giunta et al. (2009). However, we note that such model gives a statistically inadeguate representation of  the observed spectrum being the $\chi^{2}_{\nu}$=1.46 for 52 d.o.f. This motivates us to investigate  alternative spectral models.
The best fit was achieved by using an absorbed thermal black body ($\chi^{2}_{\nu}$=1.17, 52 d.o.f.)  yielding 
spectral parameters (N$_{H}$=3.4$^{+0.6}_{-0.5}$ $\times$10$^{22}$ cm$^{-2}$, kT=1.35$\pm$0.08 keV)
consistent  with previous measurements by Romano et al. (2010) from a Swift/XRT monitoring campaign 
covering an entire orbital period.  Fig. 1 shows the absorbed black body fit 
spectrum and the corresponding residuals. The unabsorbed 0.5--10 keV flux is 1.24$\times$10$^{-12}$ erg  cm$^{-2}$ s$^{-1}$ 
which translates into a X-ray luminosity of 1.3$\times$10$^{33}$ erg   s$^{-1}$ by assuming a distance of the optical counterpart of 3 kpc. 
This is the lowest X-ray state of the source  ever reported in the literature, compatible with the 3$\sigma$ upper limit 
measured by Swift/XRT in the same energy band (Romano et al. 2010). The detection of X-ray pulsations during this state (Giunta et al. 2009) strongly suggest that  it is very likely due to accretion  onto the neutron star even if at much lower rate than that during the outbursts activity. 
Above 20 keV, the lowest hard X-ray state of the source has been measured by 
IBIS/ISGRI and it is about one order of magnitude higher (Sguera et al. 2010). 
\begin{figure}[t!]
\begin{center}
\includegraphics[width=6cm,height=9cm, angle=270]{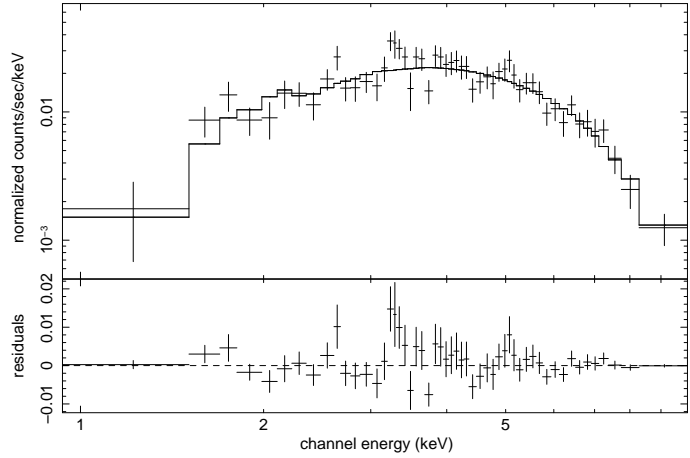}
\caption{ Data-to-model (absorbed black body) and corresponding residuals of the 0.5--10 keV EPIC-pn spectrum of IGR J18483$-$0311.}
\end{center}
\end{figure}
\begin{figure}[t!]
\begin{center}
\includegraphics[width=6cm,height=9cm, angle=270]{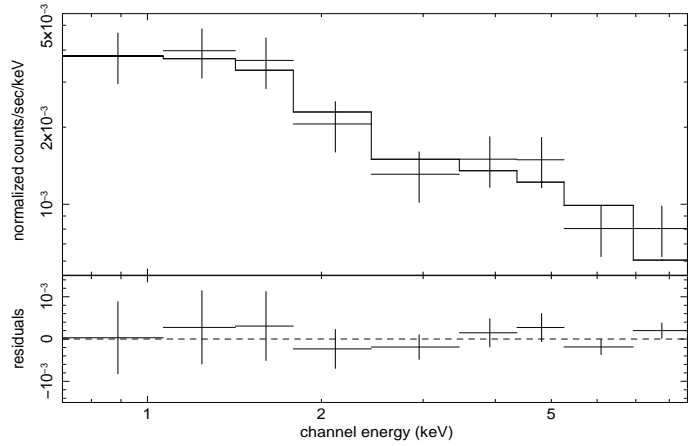}
\caption{Data-to-model (power law) and corresponding residuals of the 0.5--10 keV EPIC-pn background spectrum of IGR J18483$-$0311.}
\end{center}
\end{figure}
Although the absorbed black body model is a reasonable statistical description
of the continuum, the ratio of data to model clearly show an excess around $\sim$3.3 
keV (see Fig. 1), suggesting the presence of a  possible spectral line. Such feature is effectively 
required at a  99.4\% significance level of confidence ($\sim$3$\sigma$) according to a F-test 
($\chi^{2}_{\nu}$=0.97, 49 d.o.f), it has an energy centroid of $\sim$3.28 keV and intensity equal to
5.5$^{+1.5}_{-1.7}$ $\times$10$^{-6}$ photon cm$^{-2}$ s$^{-1}$ (uncertainties given at 68\% confidence level) 
for a significance detection of $\sim$3.5$\sigma$. 
We are aware that the F-test is not a good and proper measure of the actual significance of such 
spectral line feature (see Protassov et al. 2002), however it could give an indication and  to this aim we point out that  
the obtained low F-test probability value should make the detection of the line stable against 
mistakes in the calculation of its significance. In addition, we  also performed  a Run Test  
which can provide useful additional information. In fact, the Run Test (Barlow 1989) can be used to 
check a randomness hypothesis for a two-valued data sequence or whether a function fits well to a data set or not. 
We performed  a Run test to the residuals in Fig. 1 to test the randomness of their distribution and 
determine if there are any patterns or trends. 
As result, we obtained a chance probability equal to about 1\% that the pattern under analysis has been generated by a random process. 
Such reasonably low value provides a statistical evidence that the pattern of residuals in Fig. 1 was not generated randomly.


In order to rule out  an artificial nature of the spectral emission 
line due to instrumental effects,  we checked that the response matrix did not introduce any strong feature around 3.3 keV and 
note that no uncalibrated instrumental features or edges  are expected or known  
close to the same energy. In addition the inspection of the background spectrum, whose spectral 
shape is best fitted by a hard power law ($\chi^{2}_{\nu}$=0.6, 7 d.o.f), 
did not reveal any spectral feature around 3.3 keV as clearly visible in Fig. 2. We also note that the background
spectrum was extracted from a region of 40 arcseconds radius in the same CCD as IGR J18483$-$0311 and Fig. 3 
clearly show that no X-ray sources are located inside of it.
\begin{figure}[t!]
\begin{center}
\includegraphics[width=8cm,height=4cm]{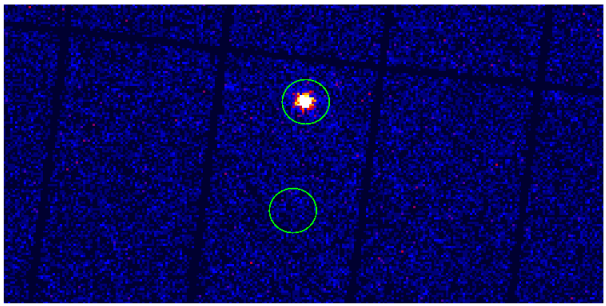}
\caption{EPIC-PN image (0.5-10 keV) of the field of IGR J18483$-$0311 during the 2006 observation. 
The source and background regions (40 arcseconds radius) considered for the analysis are superimposed.}
\end{center}
\end{figure}

\begin{figure}[t!]
\begin{center}
\includegraphics[width=7cm,height=10cm, angle=270]{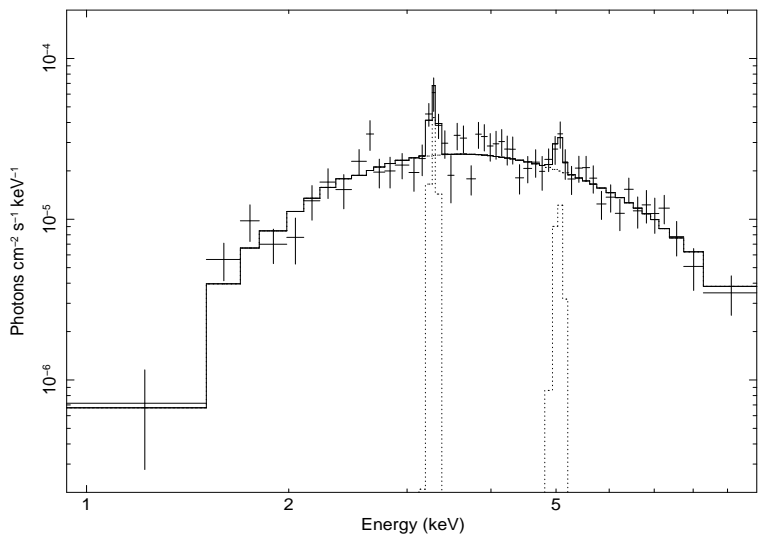}
\caption{Unfolded EPIC-pn spectrum (0.2--10 keV) best fit with an absorbed black body model plus the fundamental cyclotron line at 3.28 keV 
and its first harmonic.}
\end{center}
\end{figure}

The absence of any known or found  systematic effects  give us confidence about the non-instrumental nature of the observed spectral line. 
As for its possible physical origin, firstly we took into account the possibility that it could be due to atomic 
transition lines.  According to several atomic database (i.e. ATOMDB\footnote{http://cxc.harvard.edu/atomdb/WebGUIDE/index.html}) 
we found that two lines from highly ionized Argon are expected close to $\sim$3.3 keV (i.e. Ar XVIII at 3.32 keV and  Ar XVII at 3.14 keV). 
However, we consider unlikely a physical explanation  in terms of Ar lines because of the following reasons:
i) it would require an unexplainable high overabundance of Ar which is not obvious in the neutron star atmosphere or in its surrounding ambient, 
ii)  it would not explain why only the highly ionized Ar line is observed and no other lines are seen from 
 more abundant elements, iii) it would not justify the lack of a similar spectral feature in any of the many known accreting X-ray pulsar in HMXBs
for which much higher signal to noise X-ray spectra are available. Therefore, we explored an alternative physical explanation in terms of  
electron cyclotron emission from an X-ray pulsar accreting at a low rate, as predicted by Nelson et al. (1995,1993).
Specifically, these authors predicted the possible detection   
of electron cyclotron emission lines in the X-ray spectra of magnetized and transient X-ray pulsars during  
their low luminosity quiescent state (i.e  L$_x\leq$10$^{34}$ erg   s$^{-1}$). The energy line center  
is expected to peak at energies in the range $\sim$2--20 keV and it  should be superposed on the underlying 
soft thermal emission.  Emission-like features, similar to the ones predicted,  have been 
observed in the X-ray spectrum of a handful of transient X-ray pulsars in HMXB systems during low luminosity states (Nelson et al. 1995). 
We point out that the emission feature observed from IGR J18483$-$0311 could be similar to the one predicted by Nelson et al. (1995, 1993).
This interpretation would imply a neutron star 
magnetic field value  equal to  $\sim$2.8$\times$10$^{11}$ G 
(if the forming region is situated  far above the neutron star polar cap) or  alternatively 
$\sim$3.7$\times$10$^{11}$ G (if it is situated close to the neutron star surface, 
i.e. at the base of the accretion column,  and so affected by a gravitational redshift of z=0.3)  

In addition,  the possible detection of eventual harmonics has been also investigated. 
The first harmonic line is expected around  $\sim$5.04 keV or alternatively $\sim$6.6 keV,depending if the 
forming region is situated close or far above to the neutron star surface respectively.  
Inspection of the residuals in Fig. 1 show a very weak excess only around $\sim$5 keV, such residuals 
can be modelled with an additional gaussian line requiered at only $\sim$94\% 
significance level of confidence ($\sim$2$\sigma$) according to a simple F-test, its energy centroid and intensity are  $\sim$5.04 keV and 
2.5$^{+1.4}_{-1.2}$ $\times$10$^{-6}$ photon cm$^{-2}$ s$^{-1}$ (uncertainties given at 68\% confidence level) 
for a significance detection of $\sim$2$\sigma$. Since no atomic transition lines are expected close to $\sim$5 keV,
such possible emission feature could be related to the first harmonic of the fundamental cyclotron line.  Fig. 4 shows 
the 0.2--10 keV unfolded EPIC-pn spectrum fitted with an absorbed black body model plus the fundamental cyclotron line
and its first harmonic.

To date, IGR J18483$-$0311 is the only SFXT  which exhibits a  possible cyclotron emission feature in its X-ray spectrum.
 The lowest  X-ray luminosity state achieved during the apastron passage 
as well as the sufficiently long  observation with an appropriate and sensitive enough 
X-ray facility such as {\itshape XMM-Newton}
could have possibly favoured  the detection of the putative cyclotron emission line.
Unfortunately,  we can go no further on the above  issues because the available exposure time and statistics of the data 
prevent us from a more detailed  investigation. Longer X-ray observations of  
IGR J18483$-$0311 using {\itshape XMM-Newton}, for example, are strongly  needed 
in order to achieve a  higher signal to noise ratio. This would allow us  a much deeper investigation, 
in order to  support or reject our proposed interpretation in terms of electron cyclotron emission line. 
If firmly confirmed by a future  longer {\itshape XMM-Newton} observation, 
this would be the first detection ever of a cyclotron feature in the X-ray spectrum of a SFXT.
Its implications are very important: it will help in discriminating between different theoretical models proposed in the 
literature to physically explain the outbursts  mechanisms at work in SFXTs, i.e. involving highly 
magnetized  (B$\ge$10$^{14}$ G) or lower magnetized neutron stars (B$\sim$10$^{11}$ G).


\begin{thebibliography}{99}


\bibitem{} Barlow, R.J., 1989, The Manchester Physics Series, New York,  Wiley \\
\bibitem{} Giunta, A. et al.,  2009, MNRAS, 399, 744 \\ 
\bibitem{} Negueruela, I. et al., 2005, ESA SP-604, 165 \\
\bibitem{} Nelson, R. W. et al., 1995, ApJ, 438L, 99  \\
\bibitem{} Nelson, R. W. et al., 1993, ApJ, 418, 874  \\
\bibitem{} Protassov R. et al., 2002, ApJ, 571, 545 \\
\bibitem {} Rahoui, F., Chaty, S., 2008, A\&A, 492, 63 \\
\bibitem {} Romano, P. et al., 2010, MNRAS, 401, 1564 \\
\bibitem{} Sguera, V. et al.,  2005, A\&A, 444, 221 \\
\bibitem{} Sguera, V. et al., 2006, ApJ, 646, 452 \\ 
\bibitem {} Sguera, V. et al., 2007, A\&A, 467, 249 \\
\bibitem {} Sguera, V. et al., 2010, MNRAS, 402L, 49S \\
\bibitem {} Sidoli, L. 2009, AdSpR, 43, 1464 \\








\end{thebibliography}
\end{document}